# Indoor Positioning System Using WLAN Channel Estimates as Fingerprints for Mobile Devices

Erick Schmidt[a], David Akopian[a]

[a]The University of Texas at San Antonio, Department of Electrical Engineering, One UTSA Circle, San Antonio, TX 78249

## ABSTRACT

With the growing integration of location based services (LBS) such as GPS in mobile devices, indoor position systems (IPS) have become an important role for research. There are several IPS methods such as AOA, TOA, TDOA, which use trilateration for indoor location estimation but are generally based on line-of-sight. Other methods rely on classification such as fingerprinting which uses WLAN indoor signals. This paper re-examines the classical WLAN fingerprinting accuracy which uses received signal strength (RSS) measurements by introducing channel estimates for improvements in the classification of indoor locations. The purpose of this paper is to improve existing classification algorithms used in fingerprinting by introducing channel estimates when there are a low number of APs available. The channel impulse response, or in this case the channel estimation from the receiver, should characterize a complex indoor area which usually has multipath, thus providing a unique signature for each location which proves useful for better pattern recognition. In this experiment, channel estimates are extracted from a Software-Defined Radio (SDR) environment, thus exploiting the benefits of SDR from a NI-USRP model and LabVIEW software. Measurements are taken from a known building, and several scenarios with one and two access points (APs) are used in this experiment. Also, three granularities in distance between locations are analyzed. A Support Vector Machine (SVM) is used as the algorithm for pattern recognition of different locations based on the samples taken from RSS and channel estimation coefficients.

## 1. INTRODUCTION

Positioning systems have become evidently important and included in many applications nowadays. GPS is one example of a positioning system which uses trilateration and is very popular and available for many devices and applications [1] [2] [3], but has the limitation of working only in outdoor environments and typically requires line of sight (LOS) between transmitter and receiver, thus having a low performance in urban canyons, tunnels and other obstructed areas. Because of this limitation in GPS systems, the need for an indoor positioning system (IPS) is introduced. There are many IPS methods that are typically found in literature such as using time-of-arrival (TOA), time-difference-of-arrival (TDOA), angle-of-arrival (AOA), and combinations of these methods which also use trilateration to obtain the indoor location [4] [5], thus requiring LOS. There are also many proposals for IPS which rely on other techniques that are not based in LOS [6]. The most common and accurate method is called WLAN fingerprinting by using the received signal strength (RSS) from WLAN access points (APs) [7]. A good advantage of using WLAN fingerprinting-based methods for indoor location is that WLAN infrastructure is widely deployed in many areas such as malls and other common places, thus making it very cost-effective.

WLAN fingerprinting IPS consists of two phases: an offline and an online phase. The offline phase consists of gathering several RSS samples from many WLAN APs from a discrete number of known indoor locations. With this information, a "radio map" is generated with a desirable unique signature based on different signal strengths for each indoor location. After the offline survey, the online phase consists of measuring signal strengths and estimating an indoor location by comparing the obtained samples with the radio map stored in a database in the WLAN. RSS obtained from Wi-Fi networks is the most common variable used to assign a fingerprint on each location because of the easiness of extracting this measurement from commercial wireless network interface cards (NICs). This paper proposes to use another variable that could be extracted from the physical interaction between APs and mobile stations (MS), which is the channel estimation. Typically, indoor areas are complex to describe, thus introducing multipath. Channel estimation coefficients could prove to be a unique signature for each location because of the nature of indoor areas and because of this multipath. Although channel estimation coefficients are usually not obtained directly from proprietary hardware brands of NICs, there are other ways to extract these coefficients from the received signals by combining software-defined radio and state-of-the-art USRP receivers that operate in the ISM band. Because of the uniqueness and complexity of indoor signal propagation, it is expected that the channel will show characteristics of multipath, shadowing and other physical phenomena that should be a description found in the channel estimation for each different indoor location. Therefore, the channel impulse response, or the channel estimate in this case, should be able to describe and/or characterize each indoor location, thus making this a useful and unique signature for the classification of indoor localization via WLAN fingerprinting. If proven useful, the location accuracy will improve and provide useful for situations in which a low number of APs are available. The usefulness of indoor location based on the presence of low APs and considering channel estimation apart from the conventional RSS measurements is analyzed in this paper as an initial study. This paper will further describe the hardware and the experimental setup used to extract the channel estimation coefficients in chapter II. A description of the surveying area is also mentioned in chapter III followed by an explanation of how Support Vector Machines (SVM) are used in the experiment for the location estimation of the fingerprinting-based positioning system using RSS and channel estimates as inputs for several test scenarios in chapter IV. Finally, the results for several test scenarios are explained in chapter V followed by a conclusion.

## 2. INSTRUMENTATION

### 2.1 WLAN Standard

WLAN systems rely on the IEEE 802.11 standard which is a popular choice for wireless communications. Therefore WLANs are widely deployed in common commercial indoor areas. Thus, WLANs are a good choice for IPS. The IEEE 802.11 standard has several releases, among which the most commonly available are the following: a/b/g/n/ac/ad [8] and all of these releases are backwards compatible, i.e. an 802.11n receiver should be also able to decode packets coming with the a/b/g/n modulation formats in the physical layer. WLAN is present on the 2.4 GHz (ISM band) and 5 GHz bands. WLAN channels are usually 22 MHz wide for the 802.11b/g releases which uses DSSS modulation and 20 MHz for the 802.11a/g/n standard which uses OFDM modulation. For the ISM band, there are 14 WLAN channel frequencies available as seen in Figure 1.

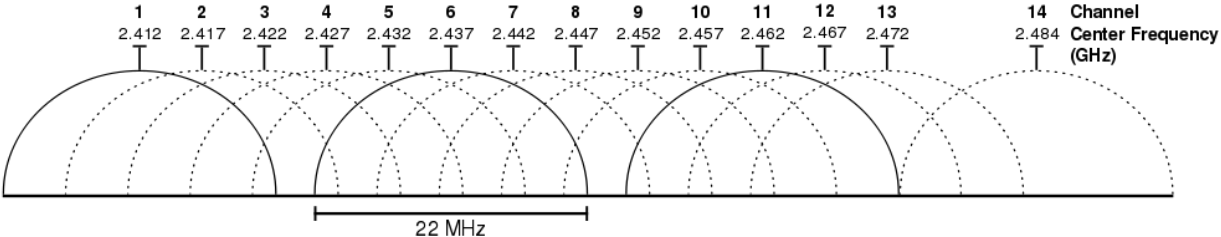

Figure 1. 2.4 GHz band WLAN Channels [9].

Because we don't need to transmit or receive actual data and/or authenticate nor connect to specific APs, we only need to extract the channel estimates and the RSS values which can be obtained from a single broadcast signal from an AP.

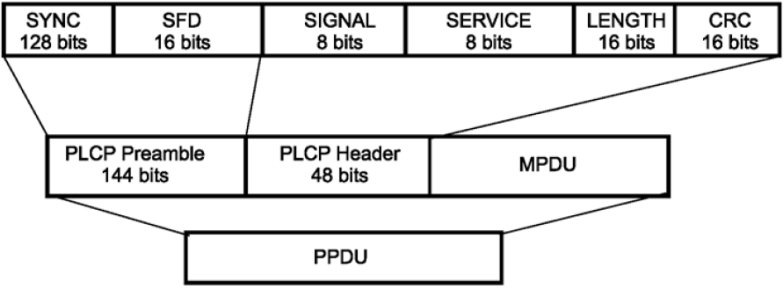

Figure 2. PLCP Frame Format [8].

These broadcast signals are carried in a specific WLAN frame and are called beacon frames. Based on [8] we usually find these beacon frames in the MAC PDU as seen in Figure 2, which is part of the PLCP PDU. This whole PPDU is then processed by the physical layer of the AP via a DSSS technique which spreads the data by an 11-length Barker code and using a DBPSK modulation. The data rate for a beacon frame is 1 Mbps. This data rate and modulation corresponds to the 802.11b release of the IEEE standard which is backwards compatible with any WLAN receiver. From the beacon frame we can extract useful information such as the SSID and the MAC address of the AP which we require for our experiment so we can distinguish between APs. At the same time we can collect physical layer information such as the channel estimation and the RSS from each successfully decoded signal. This information is contained on the MPDU which is also scrambled with the polynomial shown in Figure 3.

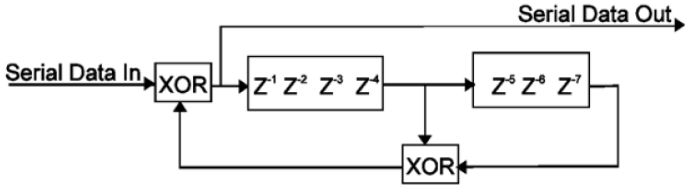

Figure 3. Data scrambler from 802.11b standard [8].

## 2.2 USRP Hardware Description

Our hardware of choice is an NI-USRP 2932 model which is capable of working on the ISM band (2.4 GHz) for WLAN. The USRP unit communicates to a laptop via a Gigabit Ethernet cable as shown in Figure 4.

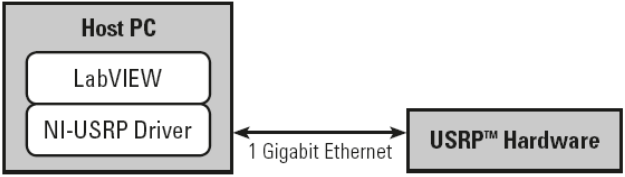

Figure 4. Interface with the USRP [10].

This radio receiver has the following configurable RX parameters:

IP Address: 192.168.10.2
Carrier Frequency (Hz): 2.412 GHz (WLAN Channel 1)
IQ Rate: 25 MHz
Active Antenna: RX1
Gain: 30 dB
Capture Time (sec): 80msec

For simplistic purposes, the carrier frequency for all the experiments was fixed to the WLAN Channel 1 which corresponds to 2412 MHz, and because the USRP device can only sample at rates multiples of 5 MHz, the selected sampling rate was 25 MHz which was later resamples to 22 MHz for compatibility with the 802.11b beacon frame. A fixed value of 30 dB was chosen for the receiver gain since it is the maximum the USRP can achieve. An 80 ms capturing time was used so that the receiver could be able to capture at least one beacon frame from the AP, which based on the IEEE 802.11 standard [8] it is transmitted every 100 ms approximately. Finally, the NI-USRP 2932 model has two reception ports, for other purposes such as MIMO and receive diversity which are out of the scope of this project. The first port named RX1 was chosen for the reception of the signal, equipped with a basic VERT 2450 antenna which is capable of reception on the 2.4-2.5 GHz band.

### 2.3 Software-Defined Radio with NI LabVIEW Environment

Software-Defined Radio is defined as a radio in which some or all of its physical layer functions are defined by software [11]. Having this definition, we implemented our own WLAN receiver in a software-defined radio environment by using LabVIEW 2013 as the software platform. LabVIEW is a platform which uses a visual programming language [12] from National Instruments and it also helps as the interface between the host PC and the USRP. We implemented the following digital communication blocks on the receiver: match filter, coarse frequency estimation and correction, code phase synchronization, equalizer, Barker code de-spreader, demodulator, and descrambler. All these digital communication blocks' descriptions were taken from the IEEE 802.11b standard recommendations [8].

### 2.4 Equalizer

For the equalizer, a Wiener-based filter was used to extract the channel estimation which was fixed to 5 channel taps. This information was extracted from each successful detection of a WLAN beacon frame and the channel estimate was collected in data logs along with the RSS from the received signal. The channel estimate is usually a complex number so the taps were further divided into real and imaginary parts, therefore having 10 components from the channel estimate and one from the RSS.

### 2.5 RSS Calculation

The RSS was calculated from its raw received signal since each WLAN NIC manufacturer uses its own measuring range, granularity, and percentage or other value such as dB [15]. We calculated the RSS by obtaining the PLCP preamble and PLCP header length which are always found as a fixed number of samples based on the standard [8]. See figure 2. The PLCP preamble is 144 bits and the PLCP header 48 bits long. The LENGTH field found on the PLCP header contains the total length of the MPDU in bits. So finally, for each correctly decoded packet, the PLCP preamble, PLCP header and MPDU length were calculated to find the total packet length. Since the location of the start of the packet is known, a simple average power was calculated from the samples at the beginning of the beacon frame, to the total length of the packet found before. With this, the sum of the raw samples divided by the total number of samples was calculated, then it was converted to dB scale to find an RSS value. Al though the RSS calculated values in dB are to be in different range, scale and granularity when compared with other commercial receivers, it is not the scope of this experiment to evaluate which scale provides the best accuracy results in classification for indoor localization. It is important to note the purpose of using RSS values which is the classification of locations based on different signal strengths, and not the actual scale used on different WLAN cards.

## 3. DATA COLLECTION

The data collection was done at the University of Texas at San Antonio on the Applied Engineering and Technology building (AET) on the 2nd floor. A rectangular area composed of 4 paths was chosen as the surveying area as seen in Figure 5.

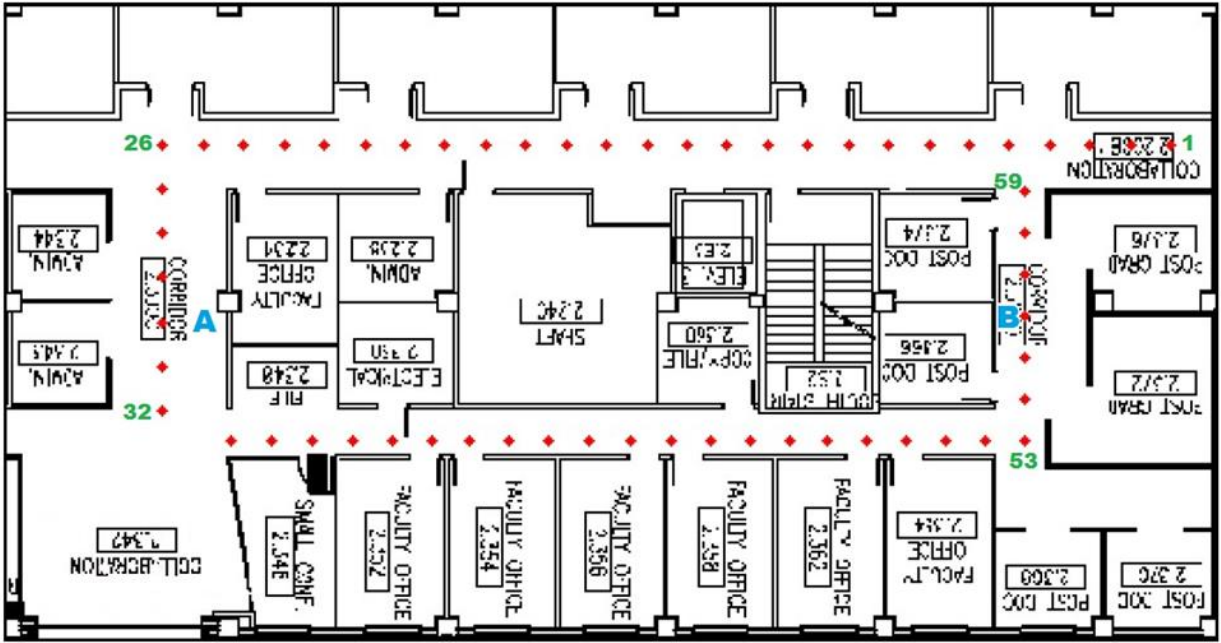

Figure 5. UTSA Campus AET Building 2nd Floor Map.

A total of 59 locations were selected which are 4 feet apart from each other. The red dots represent each location. The survey started at where location 1 is seen in a green label and ending at number 59 in green as well. Two AP were placed in location A and B with SSIDs named as TEST24 and TEST25 respectively and transmitting only on WLAN Channel 1 with a carrier frequency of 2412 MHz.

A total of 60 samples were taken on each location, using 30 samples per AP. Each sample contained the SSID, MAC-ID, RSS, and 5 real and 5 imaginary coefficients of the channel estimation as seen in Figure 6.

| SSID | MAC-ID | RSSI (dB) | W1 R | W2 R | W3 R | W4 R | W5 R | W1 I | W2 I | W3 I | W4 I | W5 I |
|---|---|---|---|---|---|---|---|---|---|---|---|---|
| TEST25 | 44-94-FC-65-F7-BA | -6.04052 | -1.90709 | 9.925677 | -11.3703 | 6.502394 | -1.54477 | 2.842852 | -2.98192 | -1.00363 | 2.875069 | -1.28599 |
| TEST24 | C0-3F-0E-90-EE-13 | -30.2728 | 43.84748 | -36.0942 | 29.07426 | -19.9224 | 3.086791 | 8.245458 | 31.2641 | -48.4419 | 45.90347 | -19.5802 |
| TEST25 | 44-94-FC-65-F7-BA | -5.44718 | -1.67688 | 4.760966 | -4.14372 | 1.829596 | -0.35705 | 0.835069 | 3.186563 | -5.33383 | 4.18504 | -1.41755 |
| TEST24 | C0-3F-0E-90-EE-13 | -31.3382 | 51.3836 | -72.0017 | 71.8827 | -57.5846 | 15.31591 | -1.60303 | 80.44502 | -87.6003 | 69.59924 | -32.0424 |

Figure 6. Sample collecting log from LabVIEW SDR WLAN 802.11b Receiver.

## 4. MATCHING ALGORITHM

### 4.1 Definition of a Support Vector Machine

Based on [6] and [7], the best matching algorithm for classification for WLAN fingerprinting using RSS is the Support Vector Machine (SVM). In this experiment, we will be using SVM algorithm for classification. The SVM is a popular method for data classification [13]. The SVM is considered as a binary classifier so based on the provided data it brings it into a higher plane and it draws a hyper plane which divides the data into either classification. The usual methodology for the classification task is to first train the SVM so it can come up with a training model, and afterwards test the SVM by providing the previously trained model. There are two categories used as inputs for either training or testing the machine: the first one is a vector of classes, and the second is a matrix of features which should correspond to each class. In this experiment we are considering the discrete locations as the classes, and the features are the RSS values of each AP as well as the channel estimates. In this experiment, a MATLAB library available online called LIBSVM from [14] will be used for the classification task in this experiment.

## 4.2 Channel estimation on the SVM

SVM takes real numbers only as inputs because of the nature of the algorithm which uses a higher plane when comparing numbers to obtain a classification match. To input complex number would have no "real" significance since SVMs are used for classification and the desired features that are used on SVMs should be unique and real for a better estimation. For this experiment, the complex numbers obtained from the channel estimation from the equalizer were further processed and only the magnitude of each channel estimation tap was used as a feature to input into the SVM for classification. Based on the measurements extracted from the SDR instrumentation used, the real and imaginary values were combined together to get a total of 5 magnitudes corresponding to each tap from the channel estimation. These were used as additional features in this experiment for a desired improvement of indoor localization. Because of the uniqueness and complexity of indoor signal propagation, it is expected that the channel will show characteristics of multipath, shadowing and other physical phenomena that should be a description found in the channel estimation for each different indoor location. In other words, the actual channel impulse response, or in this case the channel estimate, should be able to describe and/or characterize each indoor location, thus making this a useful and unique signature for an improved accuracy on indoor localization when applying classification algorithms.

## 5. RESULTS

For our first experiment we had a total of 3540 samples which correspond to 60 samples on 59 location. Each location is set 4 ft. apart as seen on Figure 5 from section III. Out of these total samples, 1770 were used for training and the other half were used for actual testing for accuracy which were inputs into the SVM. This means that on each location, 30 samples were used for training the SVM machine and 30 samples were used for location estimation. Figure 7 shows a CDF plot which shows the distance error in meters for two cases: (a) using RSS vs RSS and channel estimate magnitudes for 1 AP and (b) using RSS vs RSS and channel estimate magnitudes for 2 APs. For 1 AP case, we achieved an accuracy of 32.3% using only RSS measurements, and an accuracy of 56.4% using RSS and channel estimates. For the RSS only case with 1 AP we measured a mean distance error (50th percentile) of 3.1 m and 0.9 m for RSS and channel estimates respectively. For 2 APs case, we achieved 71.6% accuracy with a mean error of 0.7m for the RSS only case, and an accuracy of 66.3% with a mean error of 0.75m for the RSS and channel estimate case. These results are shown in Figure 7 part (a) and (b) respectively.

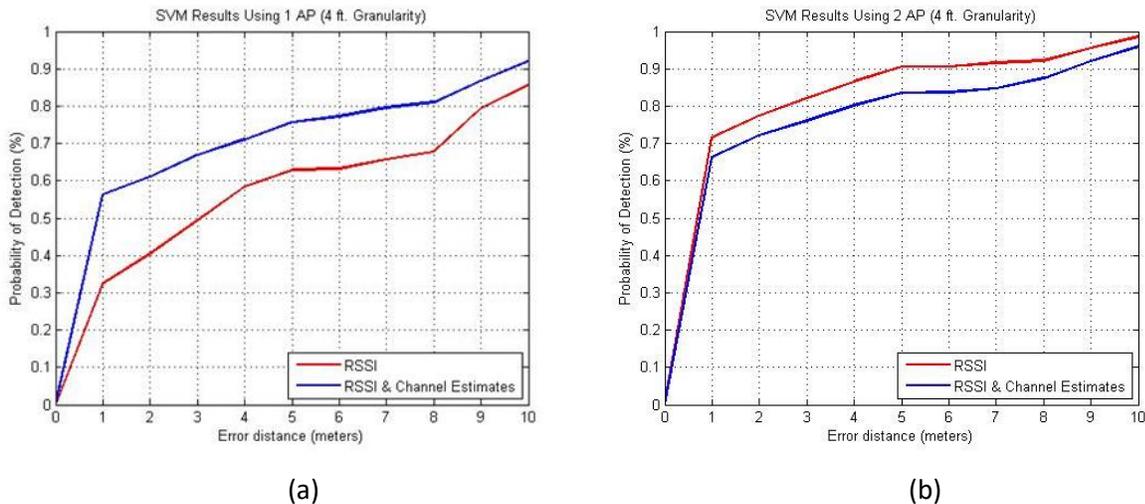

(a) (b)
Figure 7. Performance analysis of RSS vs. RSS & Channel Estimates using (a) 1 AP and (b) 2 AP with 4 ft. granularity.

The next two experiments consisted of a decimation of the same locations by a factor of 2 and 3, thus increasing the distance between samples to 8 ft. and 12 ft. respectively. This was to simulate a bigger granularity in distance to see if there are any improvements. In theory, this modification should have a more pronounced uniqueness in the fingerprint or signature for each location because of the longer distance, so it is expected to see better accuracy results by the SVM.

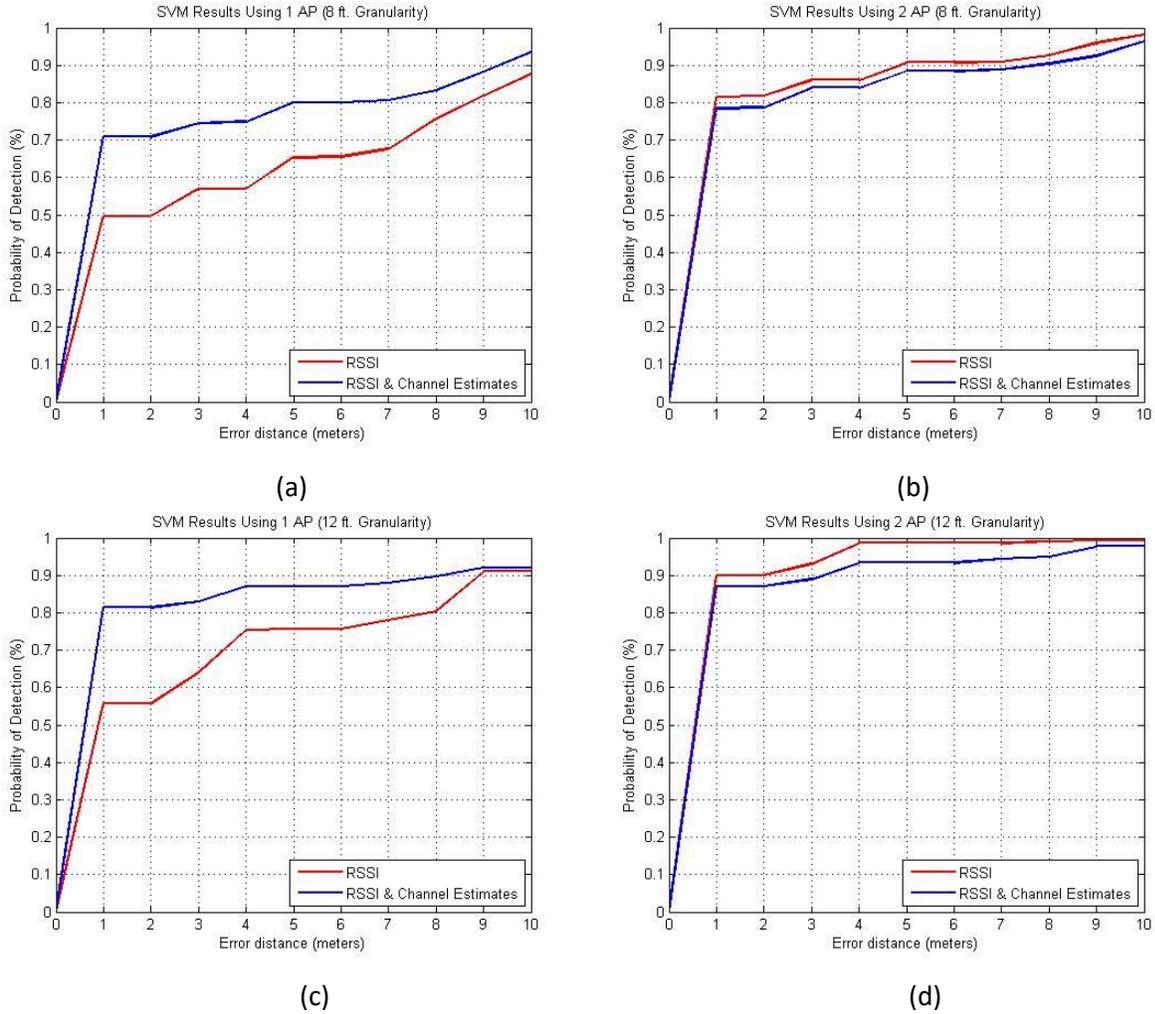

Figure 8. Performance analysis of RSS vs. RSS & Channel Estimates using (a) 1 AP and (b) 2 APs for 8 ft. and (c) 1 AP and (d) 2 APs for 12 ft.

Figure 8 show the results for experiments with 8 ft. and 12 ft. between sampled locations. The results shown in (a) with 1 AP had an accuracy of 49.6% and 70.9% for RSS only and RSS with channel estimates, with mean distance errors of 2m and 0.7m respectively. Evidently, this shows an improvement from the 4 ft. scenarios as expected. With 2 APs in (b), the accuracy was estimated as 81.6% for RSS only and 78.4% for RSS with channel estimates, with distance error mean of approximately 0.6m on both scenarios. Meanwhile, for the scenarios with 12 ft. between sampled locations, for 1 AP shown in (c) the accuracy for 1 AP with RSS only was 55.7%, and with RSS and channel estimates it was measured as 81.3%. The mean distance error for both cases was 0.9m and 0.6m respectively. Finally, for case (d) with 2 APs, the accuracy for RSS only was a very good 90% while RSS with channel estimates dropped to 87%. The respective means for RSS and RSS with channel estimates was measured as 0.56m and 0.57m which is very close. Figure 9 summarizes the results for all scenarios.

| Scenario Results for WLAN fingerprinting using RSSI and channel estimates | 4 Ft. Granularity | | | | 8 Ft. Granularity | | | | 12 Ft. Granularity | | | |
|---|---|---|---|---|---|---|---|---|---|---|---|---|
| | 1 AP | | 2 AP | | 1 AP | | 2 AP | | 1 AP | | 2 AP | |
| | RSSI only | RSSI & Channel | RSSI only | RSSI & Channel | RSSI only | RSSI & Channel | RSSI only | RSSI & Channel | RSSI only | RSSI & Channel | RSSI only | RSSI & Channel |
| Accuracy (%) | 32.3% | 56.4% | 71.6% | 66.3% | 49.6% | 70.9% | 81.6% | 78.4% | 55.7% | 81.3% | 90.0% | 87.0% |
| Mean Distance Error (meters) | 3.1 | 0.9 | 0.7 | 0.75 | 2 | 0.7 | 0.61 | 0.64 | 0.9 | 0.6 | 0.56 | 0.57 |

Figure 9. Summary results for all scenarios on WLAN location estimation via fingerprinting.

Evidently, there is an improvement in accuracy when there is only 1 AP present and channel estimate magnitudes are used. In scenarios with 2 APs, the channel estimates seem to be slightly lowering the accuracy. This could be because in this experiment, the raw values of the received channel estimates were used for classification. Further studies should be conducted to find the benefits of channel estimates and to improve the classification algorithms. Proper manipulation and post processing to the channel coefficients could yield better results. More focused research could find the actual meaning and characterization of the channel's impulse response in indoor location to find actual statistical meaning and value for better accuracy in pattern recognition. For the purpose and scope of this paper, the usage of channel estimates in conjunction with RSS and in environments with low number of APs has evidently improved matching algorithms for indoor localization estimation via fingerprinting.

## 6. CONCLUSION

This paper introduced popular IPS methodologies and explained WLAN fingerprinting-based IPS as the basis for this experiment. Apart from using RSS as the only feature for classification algorithms, channel estimates were also used on the experiment. The measurement system was explained by using an NI-USRP 2932 radio transceiver and SDR fine-tuned for decoding 802.11b WLAN beacon frames and for extracting the RSS and the channel estimates. After this, the data collection was described and the results were posted. The results showed that there is an actual improvement when there is 1 AP, proving that the channel estimation serves as a unique fingerprint since it characterizes each indoor area based on the multipath present indoors. This paper provides means for further research on the area of including channel estimation for indoor location and possible significance of the channel as a statistical approach for better accuracy in indoor localization. The channel estimates provided an improvement in the SVM matching algorithm used in this experiment, yet there is much to re-examine to maximize the accuracy of IPS by exploiting the characteristics of the channel estimate in indoor areas.